\documentclass[fonts]{icst}

\usepackage{moreverb}

\usepackage[breaklinks,colorlinks,bookmarksopen,bookmarksnumbered,linkcolor=ICSTblue,citecolor=blue,urlcolor=ICSTblue]{hyperref}
\usepackage{breakurl}
\usepackage{doi}

\newcommand\BibTeX{{\rmfamily B\kern-.05em \textsc{i\kern-.025em b}\kern-.08em
T\kern-.1667em\lower.7ex\hbox{E}\kern-.125emX}}

\def\volumeyear{2014}

\journalname{XXXXXX}
\articletype{Research Article \textbf{EAI.EU}}
 \def\volumeyear{XXXX}
 \def\volumenumber{XX}
  \def\issuemonth{XX-XX}
  \def\issuenumber{X}
  \def\articleid{XX}
  \def\copyrightholder{Patgiri\textit{ et al.}, licensed to ICST}
  \copyrightnote{This is an open access article distributed under the terms of the Creative Commons Attribution license (\url{http://creativecommons.org/licenses/by/3.0/}), which permits unlimited use, distribution and reproduction in any medium so long as the original work is properly cited.}
  \def\articledoi{10.4108/XX.X.X.XX}
\received{XXXX}
  \accepted{XXXX}
  \published{XXXX}

\begin{document}

\runningheads{Patgiri et al.}{Preventing DDoS using Bloom Filter: A Survey}

\title{Preventing DDoS using Bloom Filter: A Survey}

\author{Ripon Patgiri\affil{1}\fnoteref{1}, Sabuzima Nayak\affil{1}\fnoteref{1}, and Samir Kumar Borgohain\affil{1}\fnoteref{1}}

\address{\affilnum{1} National Institute of Technology Silchar, Assam-788010, India}

\abstract{Distributed Denial-of-Service (DDoS) is a menace for service provider and prominent issue in network security. Defeating or defending the DDoS is a prime challenge. DDoS make a service unavailable for a certain time. This phenomenon harms the service providers, and hence, loss of business revenue. Therefore, DDoS is a grand challenge to defeat. There are numerous mechanism to defend DDoS, however, this paper surveys the deployment of Bloom Filter in defending a DDoS attack. The Bloom Filter is a probabilistic data structure for membership query that returns either true or false. Bloom Filter uses tiny memory to store information of large data. Therefore, packet information is stored in Bloom Filter to defend and defeat DDoS. This paper presents a survey on DDoS defending technique using Bloom Filter.}



\fnotetext[1]{Ripon Patgiri, Corresponding author.  Email: \email{ripon@cse.nits.ac.in, sabuzimanayak@gmail.com, samir@nits.ac.in}}

\maketitle

\section{Introduction}
Distributed Denial-of-Service (DDoS) is a prominent issue in Network Security and it is extremely horrible threats for datacenters. The menace of DDoS has documented at University of Minnesota in 1991, made the system unable to serve for two days \cite{kessler}. In 2000, many media and famous companies were attacked, such as eBay, CNN, Yahoo and Amazon \cite{garber}. Usually, the DDoS attacker attacks such big companies is to make them suffer from financial losses. The losses may range to millions when they are unavailable for a few seconds \cite{bahtiyar}. In 2010, DDoS attack has done to  shut down websites such as Visa, Mastercard, PostFinance, and PayPal. Many Industries, and organizations have experienced the similar kind of peril tactics.  DDoS dissatisfies on these companies for banning the donations to WikiLeaks \cite{addley}. Moreover, there is also evidence of politically driven attack such as the most famous attack on White House Website in 2002 \cite{evans}. Also, many governmental websites are shut down during the Gezi Park revolt in Turkey \cite{Simpson}.

There are two ways to prevent the DDoS attack, namely, machine learning and Bloom Filter. This survey focuses on Bloom Filter to prevent DDoS attack. However, machine learning is intelligent to identify patterns, and to identify legal and illegal requests. But, Bloom Filter is unintelligent, hence, it cannot identify patterns and unable to differentiate legal and illegal accesses. Surprisingly, Bloom Filter is used to prevent DDoS attack. Unlike machine learning algorithm, Bloom Filter is very simple data structure that consume a tiny amount of memory. 

The Bloom Filter (BF) \cite{Bloom} is a probabilistic data structure to check the presence of an element in a set \cite{Grandi}. It is a data structure mostly used for membership filtering. Bloom Filter either returns true or false. The true result of Bloom Filter is classified into two different classes, namely, true positive and false positive. Similarly, the negative class is also classified into two different classes, particularly, false negative and true negatives. The false positive and false negative is the overhead of the filter. AS per our study, all Bloom Filter contains false positive. However, there are a few variants of Bloom Filters contain false negative. The delete operation creates a false negative issue, therefore, many Bloom filters do not support the deleting operation. But, it mostly occurs towards the saturation of the Bloom Filter.  Let, $B$ be the Bloom Filter, $S$ be the set and $K\in S$ where $K=K_1,K_2,K_3,\ldots,K_n$ and $n$ is the total number of elements. All elements of $S$ are entered into the Bloom Filter $B$. Let $K_j$ be any random query, and thus, false positive, true positive, false negative and true negative are defined as follows-
\begin{itemize}
\item True Positive: If $K_j\in B$ and $K_j\in S$, then Bloom Filter $B$ returns a true positive.
\item False Positive: If $K_j\in B$ and $K_j\not\in S$, then Bloom Filter $B$ returns a false positive.
\item False Negative: If $K_j\not\in B$ and $K_j\in S$, then Bloom Filter $B$ returns a false negative.
\item True Negative: If $K_j\not\in B$ and $K_j\not\in S$, then Bloom Filter $B$ returns a true negative.
\end{itemize}

Most of the Bloom Filter does not contain false negative, however, counting variants of Bloom Filter can encounter with false negative. A false positive/false negative is an unsolvable issue for Bloom Filter. This is an open problem that is nearly impossible to solve. However, many researchers have reduced the probability of false positive. Cuckoo Filter, for instance. Bloom Filter is an unintelligent membership filter that cannot identify patterns. On the contrary, the machine learning algorithms are intelligent to identify patterns. Both Bloom Filter and machine learning algorithm can be deployed to defeat DDoS. Tactics are different to achieve the same goal. In this study, machine learning algorithm is out of  scope.

Bloom Filter is attracting lots of attention from academia, industry, and practitioner irrespective of their research domain. Bloom Filter is used in various domains, for instance, Bioinformatics \cite{Melsted,Heo,Abyss,Chikhi}, Data-intensive computing \cite{BigTable,Lakshman}, and Networking. Thus, Bloom Filter is also used in Network Security. DDoS, for instance. It requires an excellent tactic to defend a DDoS attack using Bloom Filter. It also depends on the designing a good defense mechanism against DDoS attack. Hence, Bloom Filter is used to create a smart defender of DDoS, however, Bloom Filter is a dumb data structure. Surprisingly, most of the researcher designs a DDoS defender using Bloom Filter. It depends on the adaptation and designing the Bloom Filter to build a smart defense mechanism of the DDoS. The next section explores the deployment of Bloom Filter as defender of DDoS. Many researchers successful in defeating DDoS using Bloom Filter.

Finally, this paper presents the state-of-the-art DDoS defense mechanism using Bloom Filter. Also, DDoS attack is briefed. Moreover, DDoS defense mechanisms using Bloom Filter are explored and exposed. There are numerous research articles in DDoS defense mechanism using Bloom Filter, however, a few articles have been selected and reviewed. The DDoS defense mechanism is presented in three domains, particularly, Computer Network, Wireless Networks and Cloud Computing. DDoS attack is horrifying threats in these three domains. Moreover, this paper presents research issues and challenges in designing DDoS defense mechanism using Bloom Filter.

The article is organized as follows- Section~\ref{dos} explores on DDoS and its types.  Section~\ref{Bloom} exposes the deployment of Bloom Filter to defeat and defend the DDoS attack. Also, Section~\ref{ic} reveals the issues and challenges in defeating and defending DDoS using Bloom Filter. Finally, the article is concluded in Section~\ref{Con}.

\section{Distributed Denial-of-Service }
\label{dos}
\begin{figure*}[ht]
\centering
\includegraphics[width=0.7\textwidth,height=0.7\textwidth]{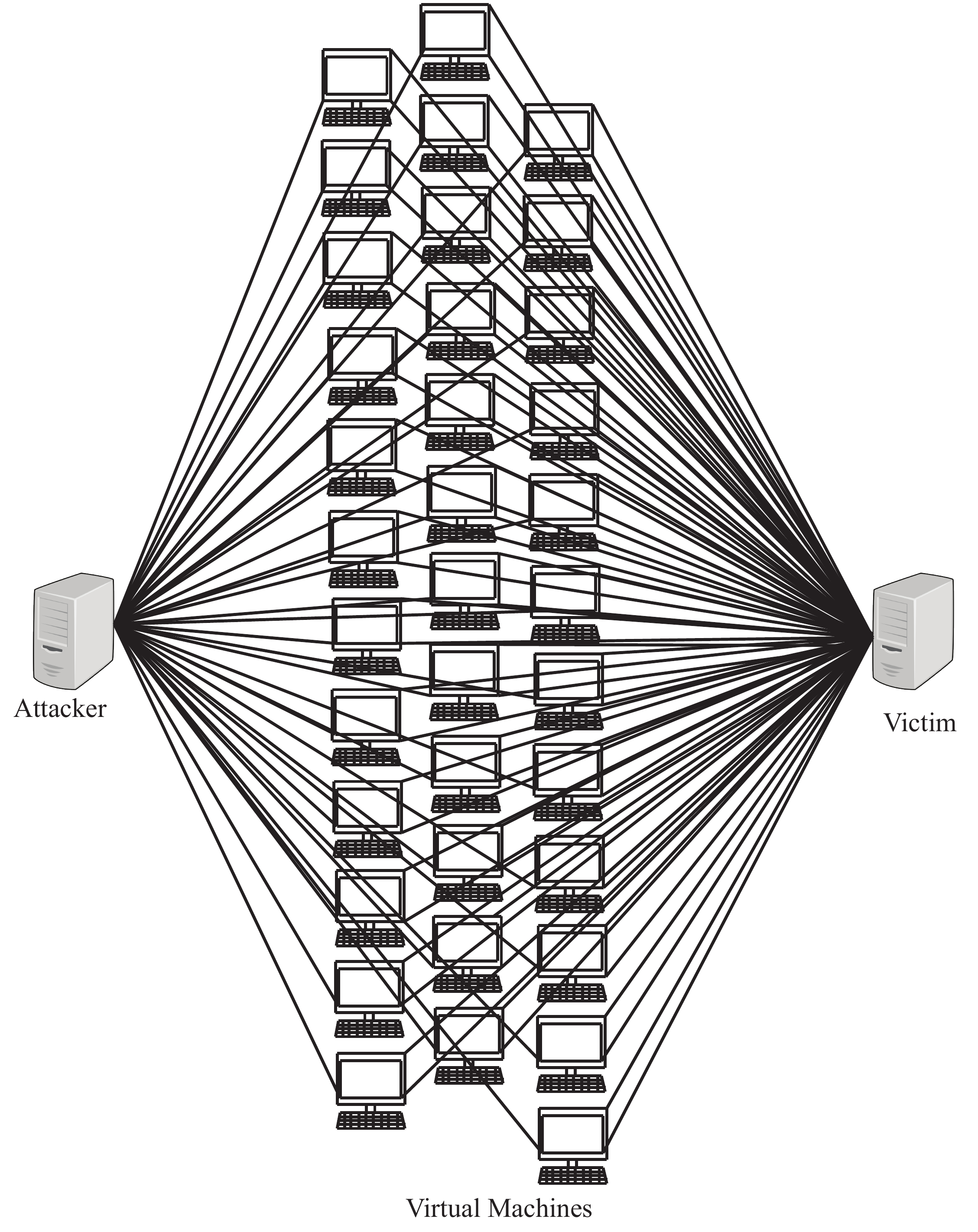}
\caption{DDoS attacker and victim. Attacker spawn several Virtual Machines to slowdown the Victim.}
\label{ddos}
\end{figure*}

In Distributed Denial-of-Service (DDoS) attack, the attacker floods the target host with millions of requests per second, which makes it unable to serve the legitimate users as depicted in Figure~\ref{ddos}. Target host is attacked from many virtual machines having different IP addresses. The traffic produced by DDoS may be in the range of hundreds of gigabits \cite{GitHub}. In 2016, the highest traffic was recorded, 1 Terabits per second \cite{Schneier}. The legitimate users are starved by these large amounts of traffic per second. DDoS attack is launched using two methods \cite{Mirkovic}- (a) Vulnerability Attack: sending some malformed packets to victim nodes. It confuses the protocol or the application running on the victim node. (b) Flooding attack: (i) disturbing the connectivity of legitimate users by exhausting router processing capacity, bandwidth, or network resources. This attack is network/transport-level flooding attacks. (ii) Disturbing the services of legitimate users by exhausting the server resources. Some server resources are CPU, sockets, disk/database bandwidth, memory, sockets, I/O bandwidth, etc. This attack is application-level flooding attacks.

DDoS attacks are often launched by Zombies or Botnet computers. The Zombies or Botnet computers are remotely controlled, well organized, and widely scattered. They concurrently and continuously send service requests to victim nodes. Usually they are recruited using Trojan horses, worms, or backdoors. They also increase their defense against detection by using spoofed IP addresses.

\subsection{DDoS Attack features}
There are many features which favor the DDoS attacker and prevents developing effective defense mechanism \cite{Peng}.
\begin{itemize}
\item A DDoS attack may generate a traffic of about 10 GB/sec. Many corporate Internet links and network security devices are unable to handle such high traffic. 
\item The attack sources comes from many distributed geographical locations. It makes very difficult for the IP traceback mechanism. 
\item As attack happened from multiple sources, the traffic generated by each source is less. This enables them to appear legitimate or as a flash crowd. Hence, filtering attacker IP addresses is difficult.
\end{itemize}

\subsection{Types}
There are five famous DDoS attacks, namely, TCP SYN Flood attack, UDP Flood attack, SQL Slammer Worm attack, DNS Amplification, and NTP attacks. 
\begin{itemize}
\item In TCP SYN Flood attack \cite{Peng}, the attacker sends a SYN packet from non-existing and not used IP addresses. The server stores the request details in memory stack and wait for confirmation from the client as per three-way handshake protocol. But the confirmation never comes. Many similar requests cause the memory to fill up, making it unable to serve the legitimate clients. There are many tools for a DDoS attack, namely, TFN, TFN2K, Stacheldraht, Shaft, and Trinity. 
\item In case of UDP flood \cite{kolahi}, a large volume of packets is sent to the victim host. The victim host becomes busy in serving those requests and thus, legitimate users are interrupted. In this attack, the victim host becomes too busy to serve other requests.
\item SQL Slammer Worm \cite{lad} sends huge request to random  hosts. If the hosts share the same MS SQL vulnerability, it gets infected and tries to infect other hosts. Often, Trinoo is used to perform the SQL Slammer Worm attack. The Slammer Worm can infect 75,000 hosts in 30 minutes \cite{moore}. 
\item In DNS Amplification attack \cite{kambourakis}, the attacker hack the DNS and create a large size resource record (RR). Then the attacker sends a request from the victim IP address a request for that RR. DNS sends the RR as response utilizing huge bandwidth for the transfer making the DNS unavailable. 
\item In NTP attack \cite{czyz}, the attacker first performs a large scale scanning to identify the vulnerable amplifiers. An amplifier is defined as a host running a protocol (e.g. NTP) which respond to a query packet with one or more packets with size greater than the query packet. When the attacker identifies such vulnerable amplifiers, sends large number of small UDP packets directly or via intermediate hosts. Such large volume of traffic saturates the bandwidth of the victim amplifier.

\end{itemize}

\subsection{DDoS defense mechanism}
The DDoS defense mechanism \cite{Peng} is classified into four broad categories, namely, attack prevention, attack detection, attack source identification, and attack reaction. 

\begin{itemize}
\item Attack prevention tries to stop the attack before the attack reaches the target host. Some examples are Ingress/Egress6 filtering at the source edge routers \cite{senie1}, Router-Based Packet Filtering \cite{Lu}. 
\item Attack detection detects the DDoS attack when it occurs. Some defense mechanisms are MULTOPS \cite{gil} and Anomaly-Based Detection \cite{garcia}. 
\item Attack source identification uses mechanisms to find the source IP address to block them. IP Traceback \cite{balyk} is the most popular mechanism to identify the attacker source IP address. 
\item Attack reaction aims to reduce or eliminate the effect of the attack. Two main approaches \cite{Peng} are taken to react to DDoS attacks, host resource management scheme and network resource management scheme.
\end{itemize}

\section{Bloom Filter and DDoS}
\label{Bloom}
Bloom Filter is a dumb data structure, however, it has a good space and time complexity which has attracted many researchers to design the DDoS defense mechanism and algorithms. In many defense mechanisms, a Bloom filter is used to store the malicious IP addresses. When a request comes, the Bloom filter is looked up for the existence of the IP address and decides whether the request is legitimate or malicious. Therefore, a scalable Bloom Filter is highly demanded to defend and defeat DDoS in a very large scale network system. There is a continuous network traffic flow in a router. The router requires a highly scalable Bloom Filter to successfully store all the information of the packet. 
 
Bloom Filter either returns true or false. There are numerous categories of Bloom Filters, however, Bloom Filter is classified into two key categories, namely, counting Bloom Filter and non-counting Bloom Filter. Counting Bloom Filters counts the number of input frequency, whereas the non-counting Bloom Filter uses fingerprints or just binary bits. Counting Bloom Filter is more scalable than non-counting Bloom Filter. However, the false positive probability of counting Bloom Filter is more than non-counting Bloom Filter.

Let $m$ be the size of Bloom Filter, and $n$ be the input size and $k$ be the number of hash functions, then the probability of number of set bits in Bloom Filter is
\begin{equation}
    \left(1-\left(1-\frac{1}{m}\right)^{nk}\right)
\end{equation}
F. Grandi \cite{Grandi} presents exact false positive probability of conventional Bloom Filter using $\delta-transformation$. Let, $X$ be the random variable represents the number of set bits and conditioning the random variable by $X=x$, then 
\begin{equation}
Pr(FP|X=x)=\left(\frac{x}{m}\right)^k    
\end{equation}
Therefore, the total false positive probability is
\begin{align}
    FPP&=\sum_{x=0}^{m} ~Pr(FP|X=x) ~Pr(X=x)\\
    FPP&=\sum_{x=0}^{m}~ \left(\frac{x}{m}\right)^k~ f(x)
\end{align}
where $f(x)$ probability mass function. F. Grandi \cite{Grandi} calculates $f(x)$ using $\delta-transformation$. Thus, the total false positive probability is
\begin{equation}\label{eq}
    FPP=\sum_{x=0}^{m}~ \left(\frac{x}{m}\right)^k~ \dbinom{m}{x} \sum_{j=0}^x \left(-1\right)^j \dbinom{x}{j} \left(\frac{x-j}{m}\right)^{nk} 
\end{equation}
Equation \eqref{eq} gives the exact false positive probability of conventional Bloom Filter. The false positive probability is an error in Bloom Filter. However, it is negligible.

\subsection{Computer network}
Kavisankar et al. \cite{kavisankar} proposed a SYN spoofing Detection and Mitigation Scheme. A Bloom Filter is used to store the legitimate IP addresses. All the requesting IP address is checked for availability every minute. These IP addresses are stored in a traffic log for 24 hours. Among these addresses the peak sample is considered for calculating the trust value. The trust value is calculated using three events, (a) reliability of network (b) behavior of node, and (c) recent status of the network. The trust value helps to store the IP Address in Bloom Filter. During DDoS attack Bloom Filter is used to determine the trusted clients.

SkyShield \cite{wang} is a sketch based defense system against application layer DDoS attack. It used two Bloom Filter, one for storing legitimate hosts and other for malicious hosts. The system does not retrieve the exact IP addresses of attacker site, and avoids intensive computation. Sketch is a data structure consist of multiple hash functions and a table. It helps in aggregating data streams of higher dimension to lesser dimension to efficiently estimate original signals. SkyShield is deployed behind a network firewall. The process has two stages mitigation and detection. During mitigation phase two Bloom Filter, namely, whitelist and blacklist. The whitelist contains the legitimate hosts. And, they are confirmed using CAPTCHA techniques. Malicious request is verified using blacklist and the information is logged. Both Bloom Filters are periodically flushed to prevent the blocking of the sites forever. The legitimate request is forwarded to detection stage. During the detection phase, anomalies are detected by exploiting the divergence between the two sketches as a signal. SkyShield is also able to detect malicious request from normal request during flash crowds.

An anti-DDoS technique\cite{tseung} is proposed which infuses self-learning to Bloom Filter. It combines both the advantages of Bloom Filter and machine learning. Initially, the packet is given to a machine learning algorithm to extract the features. The configuration is updated and given to Bloom Filter. The Bloom Filter filters and allows only the legitimate packets using the selected features.  

Halagan et al. \cite{halagan} proposed a SYN Flood attack detection and identification algorithm using the Counting Bloom Filter. A modified Counting Bloom Filter is used, called MCBF. MCBF consist of a single vector of counters. When a SYN packet is received (half-open TCP connection starts), the independent hash functions insert the IP address in MCBF and increment the counter. When an ACK packet is received (a connection is fully opened) the counter is decremented. MCBF consist of two tables. These tables have many long integer counters. First table stores the source IP addresses and second table stores the destination IP addresses. A detection algorithm called S-Orthros collects SYN packets and confirm the connection. Based on the analysis of S-Orthros, the MCBF counter is incremented or decremented. During normal connection, the MCBF remains empty. However, during the flood attack, the MCBF structure size increases very rapidly. The data stored in MCBF help to detect the type of SYN flood attack - fixed, random, subnet.

Shahsafi et al. \cite{shahsafi} proposed a Bloom filter based IP traceback method implemented in Netfilter. The Netfilter \cite{welte} is an open source framework that manipulates packets based on the Linux kernel. Netfilter used handlers called hooks for filtering of packets. The proposed IP traceback method uses a NF\_IP\_FORWARD hook to implement Bloom filter. The hook is placed in the path of the router where the packet traverse. The first router which catches the packet insert the Bloom filter to the packet based on its ID. The proposed method uses a Standard Bloom filter. In the IP header, the bloom filter is inserted into the ``Option'' field. Each router's ID is hashed to two values and inserted in the packet Bloom filter. To identify the attack, the hash values of the router given as input to the packet Bloom filter are considered as a feature in the packet. If the value is 0, the system is in normal phase and the router insert or update their hashed ID into the packet Bloom filter. However, if the value is set, the router is in trace phase. The packet Bloom filter and their hashed ID are compared to verify whether the packet has been passed through this router before or not. If the result is positive, then that packet is dropped. 

Exhaust \cite{aldwairi} is a software-based pattern matching algorithm to save the host from DDoS attack. It is based on Wu-Manber pattern matching algorithm. It uses Bloom filters to reduce the number of queries to a large hash table. Wu-Manber (WM) is a multiple pattern matching algorithm. It has two phases, preprocessing and scanning. In preprocessing phase, the minimum pattern length is found. It constructs three tables, namely, SHIFT, HASH and PREFIX. The SHIFT table contains the safe shift distance. The HASH and PREFIX tables are used for checking the full pattern against the packet which is done in the scanning phase. In scanning phase, the packet is scanned using a window of a certain size. In Exhaust, the Bloom filter is inserted between the SHIFT table and the HASH table. The HASH table content is used to program the Bloom filter. During a search of the attack signature, Exhaust slides a window over the packet and calculates the hash function which is the index of the SHIFT table. If the SHIFT table value in the index is 0, then slide the window. Otherwise, query the Bloom filter to check the presence of the string in the HASH table. If Bloom filter returns true negative, the window is shifted and the process is repeated. When the Bloom filter returns true positive, then the HASH and PREFIX tables are searched to find the exact match.

Mosharraf et al. \cite{mosharraf} proposed a responsive defense mechanism against DDoS. The model applies Bloom filter for implementing filtering close to the victim host. The model uses signature for selecting the reliable IP addresses. The signature is based on Cumulative Distribution Function of the frequency of each parameter. When the packets are received, scores are assigned based on the frequency and the signature of the selected features. The normal packets are assigned higher score. An abnormal packet is initially assigned a higher score, however, as the frequency of the packets increases the score decreases. The victim node transfers the Bloom filter to the upstream router to provide the history of the IP address. The Bloom filter is searched by the router when the packet is received. The Bloom filter helps the model by reducing the communication cost and the space overhead to store the IP address history.

\subsection{Wireless Network}
The wireless network has many issues and challenges \cite{kakkar} which makes it soft target for attacks. Some of these are Energy constraint, Bandwidth restriction, Scalability, Limited Resources, Unreliable Communication, Trust management, and Unattended Operations. In a wireless network, the devices has a limited energy and bandwidth usage. The bandwidth is susceptible to some issues, such as, interference, signal influences, and external noises. Moreover, the security methods need to be implemented for both large and small scale network. And, the limited resources in wireless devices also restrict the implementation of complex security mechanisms. The trust management is also difficult as people share a lot of personal information on wireless devices. Furthermore, many sensor nodes, remain unattended for a long time making them vulnerable to attacks. Moreover, the wireless medium is open which expose the services to eavesdroppers \cite{zhang-secure}. This leads to insecure service interaction. 

Wu et al. \cite{Wu} proposed two schemes using the counting Bloom Filter (CBF) to cramp down the SIP attack in VoLTE. SIP \cite{tang} is a Session Initiation Protocol, which is an application-layer signaling protocol. It is used for establishment, management and termination of communication sessions. SIP attack is done by attacking the victim SIP proxy server with huge SIP messages within a short period. First scheme uses authentication and CBF. Every SIP message is to be authenticated. Every SIP message carries a secret key. VoLTE carrier releases the key which is updated periodically. The key is the signature of each User Equipment (UE) in VoLTE. These signatures are stored in CBF. When an element comes IP Multimedia System (IMS) server checks its presence in CBF. The CBF also counts the number of messages to detect the occurrence of an attack. However, it occupies more memory due to which it cannot be used to detect multi-attribute flooding attack. And, the CBF has the limitation of counting at most 15 times for a signature. Second scheme utilized a PFilter inspired by CBF. All the messages are compacted to store in PFilter and the outlier is found.

A Bloom-filter based IP-CHOCK detection method \cite{verma} is proposed for the DDoS attacks in VANET. The Bloom filter is used to design a detection algorithm for making fast decision regarding filtering of vehicle attack messages. The method has three phases, namely, Detection Engine phase 1, Detection Engine phase 2, and Bloom-filter phase. First phase, checks all received traffic information. Second phase, processes the information received in the previous phase. A legitimate IP address is stored in the database. If a malicious IP address is detected, then it is sent to Decision Engine (DE). Third phase, the Bloom filter checks the DE and if a malicious IP address is verified then it raises an alarm and sends the reference link to every connected vehicle.

\subsection{Cloud Computing}
Cloud Computing is a technology based on network, and its deployment model and services makes its resources exposed to internal and external attacks \cite{masdari}. The main advantage to DDos attacker is the virtualization feature of cloud computing. Some of the DDoS attacks are VM sprawl attack, cloud-internal DoS, and VM neighbor attacks. These attacks misuse the VM migration. Currently, cloud computing technology is mostly favored, IoT and Big Data technology \cite{mtmr} are also depending more on it. The heterogeneous IoT devices makes them vulnerable \cite{wang-Hua}. If a single attacker is able to access a single device then DDoS attack can be done in the cloud to which that device is connected. Hence, cloud computing needs to focus on implementing the best security mechanisms. In Big data technology, the privacy need to be maintained while storing the data \cite{shu, zhang-efficient}. 

Xiao et al. proposed a detection system having two module detection framework \cite{xiao} against DDoS attack. This framework is designed for Software-Defined Networking (SDN). The two modules are Collector and the Detector. In the Collector module, the system scans the flow table and collects traffic flow. The flow table is obtained from the SDN network. And, the traffic flow is collected using an IP header inspection. The Detector module sniffers the entire network and collects all the packets over the network. After receiving the packets, the IP features are extracted. The extracted features are checked with the Bloom Filter to detect the abnormal flows. The Bloom filter helps in the storage of the host information. It also helps to process large traffic in high speed and stores the abnormal attack information.

Zhang et al. \cite{zhang1} proposed an efficient and robust DDoS detection model for the cloud computing. First check component monitors the UDP/TCP segment and IP flow for abnormal packets. It does IP address authentication using hop-count based filtering. Bloom filter is used for efficient address query and data storage. The Bloom filter is improved by taking a 2-Bits array. First bit is similar to Bloom filter and second bit groups stores the first pointer to the linked list. The link list stores the nodes having same $KEY$. The node stores the Source IP, hop-count and the timestamp. The $KEY$ represent the connection state of the transport layer. The improved Bloom filter is an efficient data structure for both TCP2HC and UDP2HC. It also supports efficient $KEY$ searching and robust hop-count abnormal check. The implementation of the improved Bloom filter increases the performance of the check component.

Again, Jian Zhang et al. proposed a Spark based model for identifying abnormal packets \cite{zhangJian} in Cloud network. The detection model has three components. First, the packet collector sends the packets to live input stream by libpcap. Second, the Abnormal Check component. It is implemented using pipelined task processing and Spark streaming. The Spark streaming analyzes the RDD among the DStreams. The filter, map and reduce operations helps to compute the authenticity of flow source in TCP/UDP and also checks for abnormal packets. Third, the Decision component based on Non-Parametric CUSUM. It evaluates and makes decision of any aggressive behavior. An improved Bloom filter is used for efficient lookup and storage of connection state of transport layer by HBase. The HBase stores the  hop-count related information. The improved Bloom Filter is a 2-bit array where the first bit is Bloom Filter and the second bit gives the RowID of HBASE based tables. The improved Bloom Filter provides an efficient data structure to TCP2HC and UDP2HC. 

McHale et al. \cite{mchale} proposed a model of pre-classification of packets to classify the legitimate and malicious traffic. The classification is based on the known trusted flow. Bloom filter is used to classify the packets into two queues, namely, Known Flows and Unknown flows. The flows in the queues are processed by Priority Scheduler. A Flow Cache is also included in the model to improve the classification. This cache stores flow locality within the active-flow window which helps to take action for a given flow. A unique key is extracted from the receiving packet. This key is given to Bloom filter as input. If the Bloom filter returns false positive, the packet enters the Unknown Flows queue. If no packet is present in the Known Flows queue, it is searched in the Flow Cache. If the Flow Cache also returns false response, then the packet is sent for processing through Table Selection, Flow Selection, and Action Application stages for packet classification and action-set application. During classification the Action Application stage prioritizes a flow by inserting it into Bloom filter and Flow Cache. The future packets of the flow are stored in the Known Flow queue. To reduce the false positive probability, the Bloom filter is flushed after a certain number of insertions.

\section{Key issues and challenges}
\label{ic}
\subsection{False Positive} 
False positive is a prominent issue in any kind of Bloom Filter based solution. Reducing the false positive probability is a grand challenge for the researchers. A legal user may be starved by Bloom Filter due to a false positive. Because, Bloom Filter does not know about legal or illegal accesses. Moreover, Bloom Filter is unaware of the access pattern. Hence, Bloom Filter requires an external mechanism to defeat DDoS. A designer must take utmost care of the access pattern. 

\subsection{Scalability}
DDoS defender requires a highly scalable Bloom Filter. Over a time period, Bloom Filter is filled with entries. Bloom Filter stores packet information to defend the DDoS attack. In a busy network, there are a humongous number of packet flows. Thus, all information cannot be accommodated by Bloom Filter. Therefore, scalability of Bloom Filter becomes an issue. Nevertheless, the delete operation removes the old items, however, it is not fruitful in case of DDoS defending mechanism. Most modern Bloom Filters are designed based on Flash/SSD memory to increase high scalability. For example, Forest Structured Bloom Filter \cite{Lu}, BloomStore \cite{GLu}, and BloomFlash \cite{BloomFlash}.

\subsection{Unable to send a legitimate request}
Bloom filter is used to store the malicious IP addresses. However, the attacker may have used the IP without the knowledge of the owner. Hence, the IP address is stored in the Bloom filter during the DDoS attack. And, later that IP address may never be able to send legitimate request.

\subsection{Flushing Bloom filter}
One of the solution to above problem is flushing the Bloom filter periodically. It also helps in keeping the false positive probability low. However, flushing lead to losing the information stored about the malicious IP addresses. In addition, next time the malicious IP address may not be identified as attacker site. And, allowing packets from such sites makes the host vulnerable.

\subsection{Saturated Bloom Filter} 
IP traceback defense methods use a bloom filter to store the information of the packets. Bloom filter is a data structure having less space complexity. However, storing lots of information of each packet may lead to saturation of Bloom filter very quickly. A saturated Bloom filter gives false positive results. Hence, a decision on the features of the packet that is stored by the Bloom filter need to be appropriate.

\subsection{Delete operation} 
Bloom filter is used to store the IP address. In some cases, the IP address need to be deleted from Bloom filter such as, no request from an IP address for a very long time (may be months or years). However, delete operation is not provided by all Bloom filters. Moreover, deletion of an element of Bloom filter sometime leads to false negative issue. Specially, the standard Bloom filter which is mostly used in DDoS defense mechanisms does not allow the deletion of the elements.

\section{Conclusion}
\label{Con}
 This paper presents the defending mechanism of DDoS using Bloom Filter. Bloom filter is a dump data structure, however, it helps to a great extend to tackle the DDoS attack. In many mechanisms, there is a need for a data structure that can store large number of legitimate IP address. In some other cases, the packet information need to be stored to prevent the abnormal packets reaching the host. Hence, the good space and time complexity improves security using such kind of data structures. Moreover, many types of Bloom filters are also available which can be modified to help in the prevention of the DDoS attacks. As discussed in this paper, Bloom Filter is a great data structure to prevent DDoS attack, and enhance the system performance. Moreover, Bloom Filter requires an external mechanism to define to identify the DDoS attacks. Therefore, there is a requirement for smart Bloom Filter that can learn and identify patterns for DDoS and general purpose applications. Also, a highly scalable Bloom Filter is always called for.  

\bibliographystyle{abbrv}
\bibliography{mybibfile}

\end{document}